\pdfoutput=1
\documentclass[12pt,a4paper]{article}

\usepackage{ifthen} 
\usepackage{appendix}
\newboolean{pdflatex}
\setboolean{pdflatex}{true} 

\newboolean{articletitles}
\setboolean{articletitles}{true} 

\newboolean{uprightparticles}
\setboolean{uprightparticles}{false} 

\def\paperauthors{LHCb collaboration} 
\def\paperasciititle{Measurement of the ratio of branching fractions B(Bc+ -> J/psi tau+ nu\_tau)/B(Bc+ -> J/psi mu+ nu\_mu)} 
\def\papertitle{Measurement of the ratio of branching fractions ${\mathcal{B}(\Bcp\!\!\to\! \jpsi\taup\neut)\!/\!\mathcal{B}(\Bcp\!\!\to\!\jpsi\mup\neum)}$} 
\def\paperkeywords{{High Energy Physics}, {LHCb}} 
\def\papercopyright{\the\year\ CERN for the benefit of the LHCb collaboration} 
\def\paperlicence{CC BY 4.0 licence}
\def\paperlicenceurl{https://creativecommons.org/licenses/by/4.0/}

\newif\ifEnableSectionTOCLinks
\EnableSectionTOCLinksfalse 

\usepackage[top=1in, bottom=1.25in, left=1in, right=1in]{geometry}

\usepackage{microtype}
\usepackage{lineno}  
\usepackage{xspace} 
\usepackage{caption} 

\usepackage{graphicx}  
\usepackage{color}
\usepackage{colortbl}
\graphicspath{{./figs/}} 

\usepackage{amsmath} 
\usepackage{amssymb}
\usepackage{amsfonts}
\usepackage{upgreek} 

\newcommand*\patchAmsMathEnvironmentForLineno[1]{%
\expandafter\let\csname old#1\expandafter\endcsname\csname #1\endcsname
\expandafter\let\csname oldend#1\expandafter\endcsname\csname
end#1\endcsname
 \renewenvironment{#1}%
   {\linenomath\csname old#1\endcsname}%
   {\csname oldend#1\endcsname\endlinenomath}%
}
\newcommand*\patchBothAmsMathEnvironmentsForLineno[1]{%
  \patchAmsMathEnvironmentForLineno{#1}%
  \patchAmsMathEnvironmentForLineno{#1*}%
}
\AtBeginDocument{%
\patchBothAmsMathEnvironmentsForLineno{equation}%
\patchBothAmsMathEnvironmentsForLineno{align}%
\patchBothAmsMathEnvironmentsForLineno{flalign}%
\patchBothAmsMathEnvironmentsForLineno{alignat}%
\patchBothAmsMathEnvironmentsForLineno{gather}%
\patchBothAmsMathEnvironmentsForLineno{multline}%
\patchBothAmsMathEnvironmentsForLineno{eqnarray}%
}

\usepackage[pdftex,
            pdfauthor={\paperauthors},
            pdftitle={\paperasciititle},
            pdfkeywords={\paperkeywords}]{hyperref}
\usepackage{hyperxmp}
\hypersetup{
    pdfcopyright={Copyright (C) \papercopyright},
    pdflicenseurl={\paperlicenceurl}
}

\usepackage[colorinlistoftodos,textsize=scriptsize]{todonotes}

\usepackage[bottom,flushmargin,hang,multiple]{footmisc}

\usepackage[all]{hypcap} 

\usepackage{xspace} 
\usepackage{upgreek}

\def\MagUp {\mbox{\em Mag\kern -0.05em Up}\xspace}

\ifthenelse{\boolean{uprightparticles}}%
{

 \def\Pmu         {\ensuremath{\upmu}\xspace}                 
 \def\Pnu         {\ensuremath{\upnu}\xspace}                 
                  
 \def\Ppi         {\ensuremath{\uppi}\xspace}

 \def\Ptau        {\ensuremath{\uptau}\xspace}

 \def\Ppsi        {\ensuremath{\uppsi}\xspace}

 \def\PDelta      {\ensuremath{\Delta}\xspace}                 
 \def\PXi         {\ensuremath{\Xi}\xspace}                 
 \def\PLambda     {\ensuremath{\Lambda}\xspace}                 
 \def\PSigma      {\ensuremath{\Sigma}\xspace}                 
 \def\POmega      {\ensuremath{\Omega}\xspace}                 
 \def\PUpsilon    {\ensuremath{\Upsilon}\xspace}
 \let\oldPi\Pi
 \def\PPi         {\ensuremath{\oldPi}\xspace}

 \def\PB      {\ensuremath{\mathrm{B}}\xspace}                 
 \def\PD      {\ensuremath{\mathrm{D}}\xspace}                 
 \def\PJ      {\ensuremath{\mathrm{J}}\xspace}                 
 \def\PK      {\ensuremath{\mathrm{K}}\xspace}                 
 \def\Pb      {\ensuremath{\mathrm{b}}\xspace}                 
 \def\Pc      {\ensuremath{\mathrm{c}}\xspace}

 \def\Ps      {\ensuremath{\mathrm{s}}\xspace}

 \def\thebaroffset{0.0em}
}
{

 \def\Pmu         {\ensuremath{\mu}\xspace}                 
 \def\Pnu         {\ensuremath{\nu}\xspace}                 
                  
 \def\Ppi         {\ensuremath{\pi}\xspace}

 \def\Ptau        {\ensuremath{\tau}\xspace}

 \def\Ppsi        {\ensuremath{\psi}\xspace}                 
                  
 \mathchardef\PDelta="7101
 \mathchardef\PXi="7104
 \mathchardef\PLambda="7103
 \mathchardef\PSigma="7106
 \mathchardef\POmega="710A
 \mathchardef\PUpsilon="7107
 \mathchardef\PPi="7105
 \def\PB      {\ensuremath{B}\xspace}                 
 \def\PD      {\ensuremath{D}\xspace}                 
 \def\PJ      {\ensuremath{J}\xspace}                 
 \def\PK      {\ensuremath{K}\xspace}                 
 \def\Pb      {\ensuremath{b}\xspace}                 
 \def\Pc      {\ensuremath{c}\xspace}

 \def\Ps      {\ensuremath{s}\xspace}

 \def\thebaroffset{0.18em}
}
\newcommand{\offsetoverline}[2][\thebaroffset]{\kern #1\overline{\kern -#1 #2}}%

\makeatletter
\ifcase \@ptsize \relax
  \newcommand{\miniscule}{\@setfontsize\miniscule{4}{5}}
\or
  \newcommand{\miniscule}{\@setfontsize\miniscule{5}{6}}
\or
  \newcommand{\miniscule}{\@setfontsize\miniscule{5}{6}}
\fi
\makeatother

\DeclareRobustCommand{\optbar}[1]{\shortstack{{\miniscule (\rule[.5ex]{1.25em}{.18mm})}
  \\ [-.7ex] $#1$}}

\def\mup        {{\ensuremath{\Pmu^+}}\xspace}
\def\mun        {{\ensuremath{\Pmu^-}}\xspace} 

\def\taup       {{\ensuremath{\Ptau^+}}\xspace}
\def\taum       {{\ensuremath{\Ptau^-}}\xspace}

\def\lepton     {{\ensuremath{\ell}}\xspace}

\def\neu        {{\ensuremath{\Pnu}}\xspace}
\def\neub       {{\ensuremath{\overline{\Pnu}}}\xspace}
\def\neum       {{\ensuremath{\neu_\mu}}\xspace}
\def\neumb      {{\ensuremath{\neub_\mu}}\xspace}
\def\neut       {{\ensuremath{\neu_\tau}}\xspace}
\def\neutb      {{\ensuremath{\neub_\tau}}\xspace}

\def\squark    {{\ensuremath{\Ps}}\xspace}

\def\cquark    {{\ensuremath{\Pc}}\xspace}

\def\bquark    {{\ensuremath{\Pb}}\xspace}

\def\pion   {{\ensuremath{\Ppi}}\xspace}

\def\pip    {{\ensuremath{\pion^+}}\xspace}

\def\kaon    {{\ensuremath{\PK}}\xspace}

\def\KorKbar {\kern \thebaroffset\optbar{\kern -\thebaroffset \PK}{}\xspace}

\def\Km      {{\ensuremath{\kaon^-}}\xspace}

\def\D       {{\ensuremath{\PD}}\xspace}

\def\DorDbar {\kern \thebaroffset\optbar{\kern -\thebaroffset \PD}\xspace}
\def\Dz      {{\ensuremath{\D^0}}\xspace}

\def\Dp      {{\ensuremath{\D^+}}\xspace}
\def\Dm      {{\ensuremath{\D^-}}\xspace}

\def\DpDm    {\ensuremath{\Dp {\kern -0.16em \Dm}}\xspace}

\def\Dstarp  {{\ensuremath{\D^{*+}}}\xspace}

\def\Ds      {{\ensuremath{\D^+_\squark}}\xspace}
\def\Dsp     {{\ensuremath{\D^+_\squark}}\xspace}

\def\B       {{\ensuremath{\PB}}\xspace}
\def\Bbar    {{\ensuremath{\offsetoverline{\PB}}}\xspace}
\def\Bb      {{\ensuremath{\Bbar}}\xspace}
\def\BorBbar {\kern \thebaroffset\optbar{\kern -\thebaroffset \PB}\xspace}

\def\Bd      {{\ensuremath{\B^0}}\xspace}

\def\BdorBdbar {\kern \thebaroffset\optbar{\kern -\thebaroffset \Bd}\xspace}

\def\Bs      {{\ensuremath{\B^0_\squark}}\xspace}

\def\BsorBsbar {\kern \thebaroffset\optbar{\kern -\thebaroffset \Bs}\xspace}

\def\Bcp     {{\ensuremath{\B_\cquark^+}}\xspace}

\def\jpsi     {{\ensuremath{{\PJ\mskip -3mu/\mskip -2mu\Ppsi}}}\xspace}
\def\psitwos  {{\ensuremath{\Ppsi{(2S)}}}\xspace}

\def\Y#1S{\ensuremath{\PUpsilon{(#1S)}}\xspace}

\def\LorLbar     {\kern \thebaroffset\optbar{\kern -\thebaroffset \PLambda}\xspace}

\def\BF         {{\ensuremath{\mathcal{B}}}\xspace}

\newcommand{\decay}[2]{\ensuremath{\mathinner{#1\!\to #2}}\xspace}

\def\to                 {\ensuremath{\rightarrow}\xspace}

\def\AT#1     {\ensuremath{A_{\mathrm{T}}^{#1}}\xspace}           

\def\C#1      {\ensuremath{\mathcal{C}_{#1}}\xspace}                       
\def\Cp#1     {\ensuremath{\mathcal{C}_{#1}^{'}}\xspace}                    
\def\Ceff#1   {\ensuremath{\mathcal{C}_{#1}^{\mathrm{(eff)}}}\xspace}        
\def\Cpeff#1  {\ensuremath{\mathcal{C}_{#1}^{'\mathrm{(eff)}}}\xspace}       
\def\Ope#1    {\ensuremath{\mathcal{O}_{#1}}\xspace}                       
\def\Opep#1   {\ensuremath{\mathcal{O}_{#1}^{'}}\xspace}                    

\newcommand{\aunit}[1]{\ensuremath{\text{\,#1}}}       

\newcommand{\tev}{\aunit{Te\kern -0.1em V}\xspace}
\newcommand{\gev}{\aunit{Ge\kern -0.1em V}\xspace}
\newcommand{\mev}{\aunit{Me\kern -0.1em V}\xspace}
\newcommand{\kev}{\aunit{ke\kern -0.1em V}\xspace}
\newcommand{\ev}{\aunit{e\kern -0.1em V}\xspace}
 
\newcommand{\mevc}{\ensuremath{\aunit{Me\kern -0.1em V\!/}c}\xspace}
\newcommand{\gevc}{\ensuremath{\aunit{Ge\kern -0.1em V\!/}c}\xspace}
\newcommand{\mevcc}{\ensuremath{\aunit{Me\kern -0.1em V\!/}c^2}\xspace}
\newcommand{\gevcc}{\ensuremath{\aunit{Ge\kern -0.1em V\!/}c^2}\xspace}

\def\mm   {\aunit{mm}\xspace}

\def\fb   {\ensuremath{\aunit{fb}}\xspace}
\def\invfb   {\ensuremath{\fb^{-1}}\xspace}

\def\ps   {\ensuremath{\aunit{ps}}\xspace}

\newcommand{\stat}{\aunit{(stat)}\xspace}
\newcommand{\syst}{\aunit{(syst)}\xspace}

\def\gsim{{~\raise.15em\hbox{$>$}\kern-.85em
          \lower.35em\hbox{$\sim$}~}\xspace}
\def\lsim{{~\raise.15em\hbox{$<$}\kern-.85em
          \lower.35em\hbox{$\sim$}~}\xspace}

\def\pt         {\ensuremath{p_{\mathrm{T}}}\xspace}

\def\tell1  {TELL1\xspace}
\def\ukl1   {UKL1\xspace}

\newcommand{\lhcborcid}[1]{\href{https://orcid.org/#1}{\hspace*{0.1em}\raisebox{-0.45ex}{\includegraphics[width=1em]{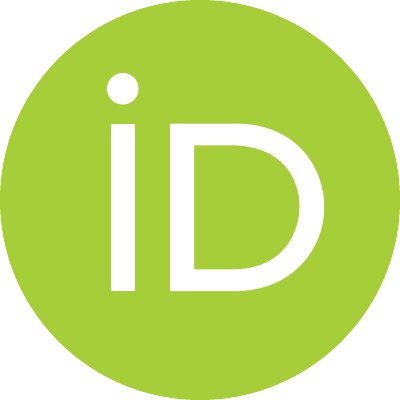}}}}

\hypersetup{
  colorlinks   = true, 
  urlcolor     = blue, 
  linkcolor    = blue, 
  citecolor    = red   
}

\ifEnableSectionTOCLinks
    \usepackage[explicit]{titlesec} 
    
    \let\oldcontentsline\contentsline
    \renewcommand

    \titleformat{\section}{\normalfont\Large\bf}{\hyperlink{tocsection.\thesection}{{\thesection} \parbox[t]{\dimexpr\textwidth-1pc}{#1}}}{1pc}{}

    \titleformat{\subsection}{\normalfont\bf}{\hyperlink{tocsubsection.\thesubsection}{{\thesubsection} \parbox[t]{\dimexpr\textwidth-1pc}{#1}}}{1pc}{}

    \titleformat{name=\section,numberless}[display]{}{}{0pt}{\normalfont\Huge\bfseries #1}
\fi

\usepackage{cite} 
\usepackage{LHCb/mciteplus}

\usepackage{longtable} 

\def\FinalVal{\ensuremath{0.51 \pm 0.12 \stat \pm 0.08 \syst}\xspace}
\usepackage{float}

\begin{document}

\renewcommand{\thefootnote}{\fnsymbol{footnote}}
\setcounter{footnote}{1}


\begin{titlepage}
\pagenumbering{roman}

\vspace*{-1.5cm}
\centerline{\large EUROPEAN ORGANIZATION FOR NUCLEAR RESEARCH (CERN)}
\vspace*{1.5cm}
\noindent
\begin{tabular*}{\linewidth}{lc@{\extracolsep{\fill}}r@{\extracolsep{0pt}}}
\ifthenelse{\boolean{pdflatex}}
{\vspace*{-1.5cm}\mbox{\!\!\!\includegraphics[width=.14\textwidth]{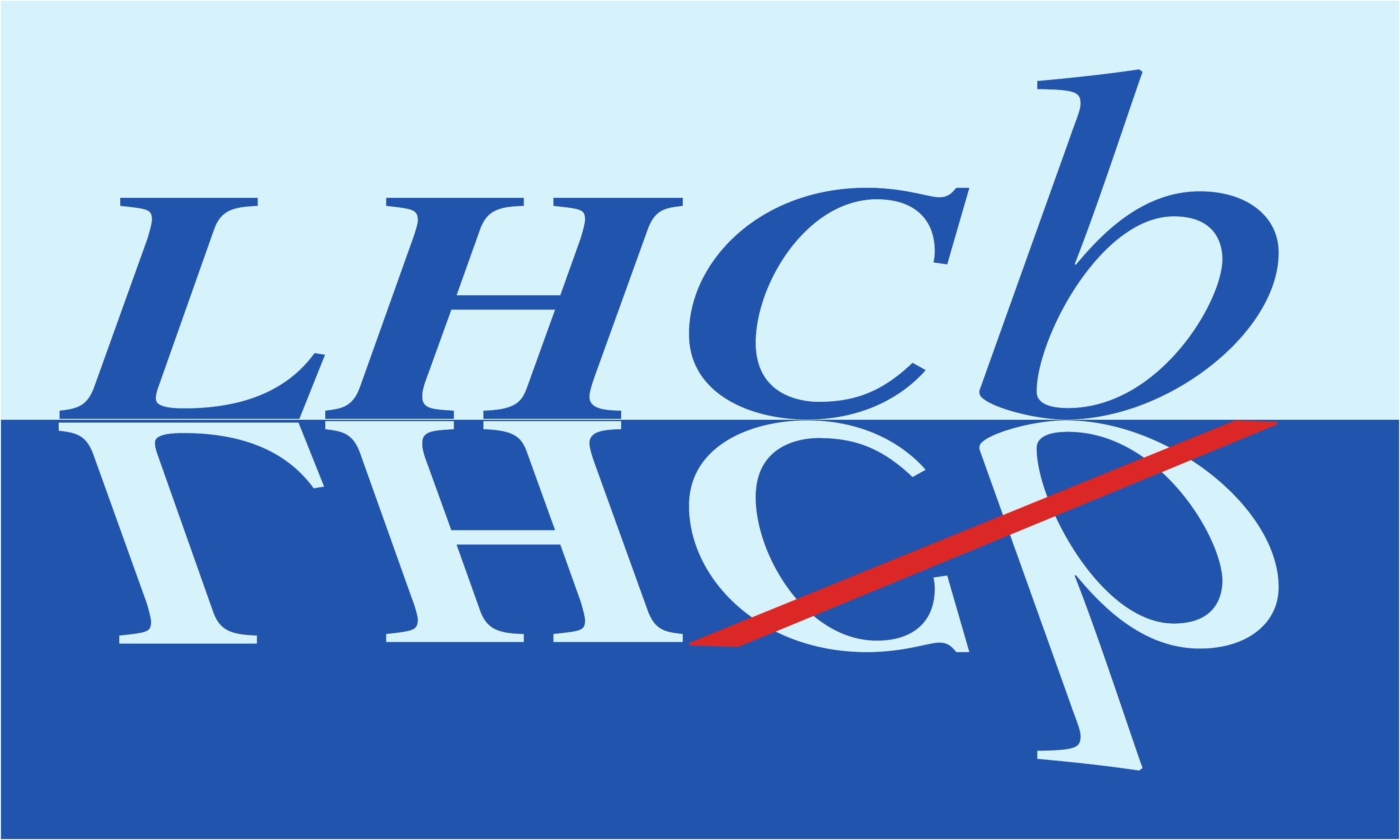}} & &}%
{\vspace*{-1.2cm}\mbox{\!\!\!\includegraphics[width=.12\textwidth]{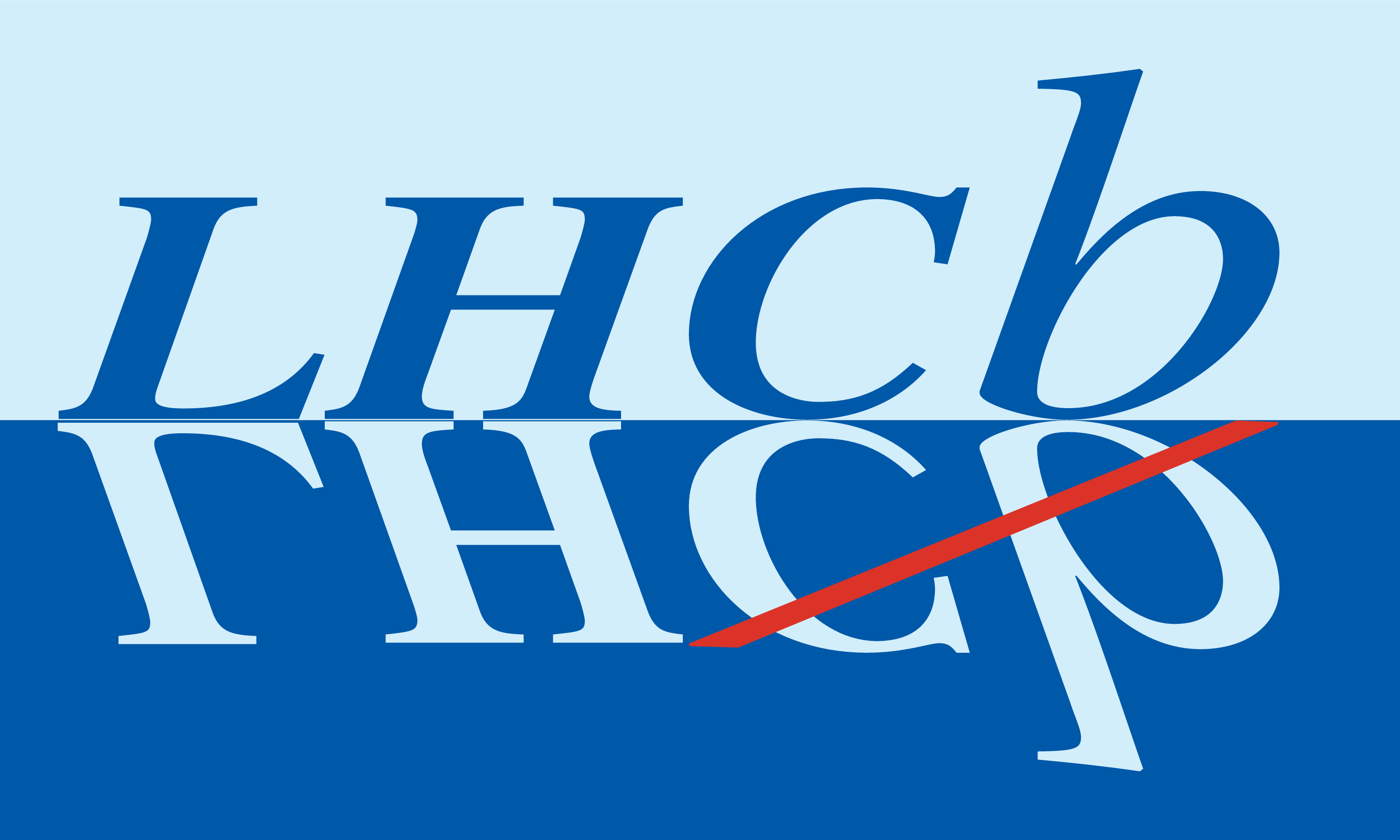}} & &}%
\\
 & & CERN-EP-2026-152 \\  
 & & LHCb-PAPER-2026-018 \\  
 & & \today \\ 
\end{tabular*}

\vspace*{4.0cm}

{\normalfont\bfseries\boldmath\huge
\begin{center}
  \papertitle 
\end{center}
}

\vspace*{2.0cm}

\begin{center}
\paperauthors\footnote{Authors are listed at the end of this Letter.}
\end{center}

\vspace{\fill}

\begin{abstract}
  \noindent
  A measurement of the ratio of semileptonic branching fractions $\mathcal{R}(\jpsi )$, defined as \mbox{$\mathcal{R}(\jpsi ) \equiv {\mathcal{B}(B_c^+ \to \jpsi \tau^+ \nu_{\tau})}/{\mathcal{B}(B_c^+ \to \jpsi \mu^+ \nu_{\mu})}$}, is reported using a sample of proton-proton collision data corresponding to an integrated luminosity of 5.4~fb$^{-1}$ recorded by the LHCb experiment in 2016--2018 at a center-of-mass energy of 13\tev. The measured value is found to be $\mathcal{R}(\jpsi) = 0.51 \pm 0.12\stat \pm 0.08\syst$, which is within 1.8 standard deviations of the predictions from the Standard Model assuming lepton flavor universality. 
  
\end{abstract}

\vspace*{2.0cm}

\begin{center}
  Published in
  Phys.~Rev.~Lett. 137 (2026) 141801 
\end{center}

\vspace{\fill}

{\footnotesize 
\centerline{\copyright~\papercopyright. \href{\paperlicenceurl}{\paperlicence}.}}
\vspace*{2mm}

\end{titlepage}


\newpage
\setcounter{page}{2}
\mbox{~}
%
%
%
%


\renewcommand{\thefootnote}{\arabic{footnote}}
\setcounter{footnote}{0}


\cleardoublepage


\pagestyle{plain} 
\setcounter{page}{1}
\pagenumbering{arabic}


Semileptonic decays of $\bquark$ hadrons allow for precise tests of lepton flavor universality (LFU), a fundamental feature of the Standard Model (SM) that entails equal couplings for all charged leptons to the gauge bosons. Measurements of the ratios $\mathcal{R}(D)$ and $\mathcal{R}(D^*)$, where $\mathcal{R}(D^{(*)}) \equiv \BF (\decay{\Bb}{D^{(*)} \taum \neutb})/\BF (\decay{\Bb}{D^{(*)} \mun \neumb})$ 
show an excess of semitauonic decays, representing a persistent tension with the SM~\cite{HFLAV:2024ctg}. Here and throughout this Letter, charge-conjugate processes are implied. The data recorded by the LHC experiments have allowed the extension of these studies to other $b$-hadron species, notably the $\Bcp$ meson.
Using proton-proton ($pp$) collision data acquired with the LHCb detector during 2011--2012 (LHC Run 1), corresponding to an integrated luminosity of $3\invfb$ at center-of-mass energies of 7 and 8\tev, the LHCb collaboration first reported a measurement of the ratio of branching fractions
\begin{equation*}
    \mathcal{R}(\jpsi) \equiv \frac{\mathcal{B}(\decay{\Bcp}{\jpsi \taup \neut})}{\mathcal{B}(\decay{\Bcp}{\jpsi \mup \neum})},
\end{equation*}
with a result of $0.71 \pm 0.17 \stat \pm 0.18 \syst$\cite{LHCb-PAPER-2017-035}. This is within two standard deviations of the SM predictions available at the time, with central values ranging from 0.25 to 0.28~\cite{Anisimov:1998uk,Kiselev,Ivanov:2006ni,Hernandez:2006gt}. Two subsequent measurements of this observable have also been reported by the CMS collaboration with a combined value of $\mathcal{R}(\jpsi)=0.49\pm0.26$~\cite{CMS:2024seh,CMS-combined-result}, consistent with the SM prediction. Major progress has also been achieved on the theoretical front,  with lattice QCD (LQCD) calculations of the $\decay{B_c^+}{\jpsi \ell^+ \nu_{\ell}}$ form factors. With these inputs, the value of $\mathcal{R}(\jpsi)$, assuming LFU, is predicted to be $0.2597 \pm 0.0027$\cite{Harrison:2025}. The improved theoretical input on the form factors, in particular, is critical to the reduction of the corresponding systematic uncertainties in these types of experimental measurements.

In this Letter, a measurement of $\mathcal{R}(\jpsi)$ is presented using $pp$ collision data acquired with the LHCb detector during 2016--2018 (LHC Run 2) at a center-of-mass energy of 13\tev, and corresponding to an integrated luminosity of $5.4\invfb$. Relative to the previous LHCb measurement, the analyzed $b$-hadron sample size is larger by a factor of four owing to a nearly twofold increase in the $b$-hadron production cross section at the higher $pp$ collision energy~\cite{LHCb-PAPER-2016-031}.

The decays $\decay{\Bcp}{\jpsi \tau^+ \neut}$ and $\decay{\Bcp}{\jpsi \mup \neum}$, referred to in this Letter as the signal and normalization decay channels, respectively, are reconstructed using $\decay{\jpsi}{\mup \mun}$ and the purely leptonic decay $\decay{\tau^+}{\mup \neum \overline{\nu}_{\tau}}$, leading to a visible signature of $(\mup\mun)\mup$ for both channels. The muon candidate not originating from the \jpsi decay is hereafter referred to as the unpaired muon. The signal and normalization channels are distinguished from each other and from background processes by exploiting differences in their kinematic distributions (see Supplemental Material for distributions~\cite{ref:SuppMat}). Their relative yields are estimated using a multidimensional fit to the data, with each component represented by a template distribution derived either from simulation or control data samples.

The LHCb detector is a single-arm forward spectrometer covering the pseudorapidity range $2<\eta<5$, described in detail in Refs.~\cite{LHCb-DP-2008-001,LHCb-DP-2014-002}. Critically for this analysis, muons are identified by a system composed of alternating layers of iron and multiwire proportional chambers~\cite{LHCb-DP-2013-001}. The online event selection is performed by a trigger~\cite{LHCb-DP-2012-004}, which consists of a hardware stage, based on information from the calorimeter and muon systems, followed by a software stage, which applies a full event reconstruction.

Events containing $\jpsi \mup$ candidates are first selected by the LHCb hardware dimuon trigger. The software trigger then identifies $\decay{\jpsi}{\mup \mun}$ candidates using oppositely charged good-quality tracks that pass muon identification requirements. These muons must have $p>6\gevc$ and $\pt>200\mevc$, with $p$ being the momentum and $\pt$ its component transverse to the beam line. The $\jpsi$ is required to have $\pt> 2\gevc$, and must be significantly detached from the primary collision vertex~(PV) to suppress background from prompt \jpsi mesons produced directly at the collision point.

Further selections are applied offline in order to reduce background contributions. The number of tracks in the event is required to be fewer than 600. 

The muons selected to form the $\jpsi$ candidate must have an impact parameter $\chi^2$ greater than 4. The $\jpsi$ candidate is required to have an invariant mass within $55\mevcc$ of the known $\jpsi$ mass~\cite{PDG2024} and form a good-quality vertex. 
The unpaired muon candidate is required to have momentum $3<p<100\gevc$ and $\pt>750\mevc$ and it is required to be inconsistent with originating from a PV. 
To remove candidates where multiple opposite-sign muon pairs are consistent with originating from a $\jpsi$ decay, 
the combinations formed from the unpaired muon and the opposite-sign muon from the \jpsi candidate are rejected if their invariant mass is within $50\mevcc$ of the known $\jpsi$ mass. The angle between the $\jpsi$ and unpaired muon candidates in the plane transverse to the beam axis is required to have a cosine greater than $-0.8$, as decay products of a single \bquark hadron with nonzero transverse momentum are unlikely to be back-to-back in the transverse plane. The invariant mass of the $\jpsi\mup$ candidate is required to be greater than $3.2\gevcc$, the sum of the $\jpsi$ and $\mu$ masses, but less than $6.4\gevcc$, which is just above the $\Bcp$ mass.

To reduce specific background processes, additional selections are imposed. 
Stringent particle identification selections are required to suppress candidates in which a hadron is misidentified as the unpaired muon, representing the leading background component in the data set. Standard LHCb muon identification algorithms are applied~\cite{LHCb-PROC-2011-008}, together with the requirement that hits in the muon chamber must not be shared with other tracks in the event. Additionally, the boosted decision tree algorithm developed in Ref.~\cite{LHCb-PAPER-2022-039} is trained to provide an approximately uniform identification efficiency for real muons as a function of $p$ and \pt. The combined selection has an efficiency of about 50\% for real unpaired muons.

Several backgrounds arise from partially reconstructed decays $\decay{B_c^+}{\jpsi H_c X}$ and $\decay{B_c^+}{\jpsi (\mathrm{n}\pi) \mup \neum}$, where $H_c$ refers to a charmed hadron, $X$ refers to zero or more light hadrons, and $\mathrm{n}\geq 2$. These decays are typically associated with extraneous charged particles nearby, thus their contribution to the selected sample is reduced by requiring that the $\jpsi \mup$ candidate is isolated from other tracks in the event. This is achieved through the use of the BDT algorithm described in Ref.~\cite{LHCb-PAPER-2015-025} that assigns a score evaluating the likelihood of the track originating from the $B_c^+$ candidate rather than from the rest of the event.

The large $\mu$--$\tau$ mass difference and the additional neutrinos in the signal decay result in distinct kinematic distributions relative to the normalization mode. The kinematic quantities used to discern the composition of the selected sample are the energy of the unpaired muon in the $B_c^+$ rest frame, $E^*_{\mu}$; the squared missing mass, $m_{\text{miss}}^2 \equiv (p_{B_c^+} - p_{\jpsi} - p_{\mup})^2$; the decay time of the \Bcp candidate $\tau$; and the squared four-momentum transfer to the lepton system, $q^2 \equiv (p_{B_c^+} - p_{\jpsi})^2$. Here $p_{\jpsi}$ and $p_{\mup}$ are the measured four-momenta of the $\jpsi$ meson and the unpaired muon, respectively, while $p_{B_c^+}$ denotes the estimated four-momentum of the $B_c^+$ meson using the approximation developed in Ref.~\cite{LHCb-PAPER-2015-025}, which uses the flight direction of the $B_c^+$ candidate and the invariant mass and momenta of its visible decay products. The same approximation is used to determine the frame in which $E^*_{\mu}$ and $\tau$ are computed.

The contribution of each process is estimated by fitting to the data a sum of three-dimensional histogram templates binned in $m_{\text{miss}}^2$, $\tau$, and a categorical variable $Z$, which maps intervals from the $(E_{\mu}^*, q^2)$ plane to the integers 0 to 7. The 8 values are defined in 4 $E_{\mu}^*$ intervals with edges $[0, 0.68,1.15,1.64,3.18]\gev$, with 0 to 3 satisfying $q^2<7.15\gev^2/c^4$ and 4 to 7 satisfying $q^2>7.15\gev^2/c^4$. The decay time of the $B_c^+$ candidate is binned in five intervals with edges $[0.45, 0.68,1.06,1.44,1.82,2.20]\ps$. This binning scheme follows that used in the Run 1 result~\cite{LHCb-PAPER-2017-035}, and accounts for correlations among the three variables.

The background processes mimicking the visible signatures of the signal and normalization channels include the feed-down from the decays $\decay{B_c^+}{\psitwos \mup\neum}$, $\decay{B_c^+}{\psitwos \tau^+\neut}$, $\decay{B_c^+}{\chi_{c1}\mup\neum}$, $\decay{B_c^+}{\chi_{c2}\mup\neum}$, as well as backgrounds from $\decay{B_c^+}{\jpsi H_c X}$. These contributions are modeled using simulation. The expected contributions from $\decay{B_c^+}{\chi_{c(1,2)}\tau^+\neut}$ are kinematically similar to $\decay{B_c^+}{{\psitwos}\tau^+\nu_{\tau}}$ and are within the uncertainty of that yield, so they are not explicitly included in the fit. Feed-down background yields are constrained with respect to their theoretical predictions~\cite{Ebert,Kiselev,Wang:2009mi} during the minimization of the log-likelihood, using a log-normal prior with a factor-of-two width on the corresponding yield. The ratio $\mathcal{R}(\psitwos) = \BF (\decay{B_c^+}{\psitwos \tau^+ \neut}) / \BF (\decay{B_c^+}{\psitwos \mup \neum})$ is loosely constrained around the predicted SM value of $8.5\%$~\cite{Kiselev}.

The contribution of the decays $\decay{B_c^+}{\jpsi H_c X}$ is modeled using a cocktail of two-body, quasi-two-body with higher \Ds resonances, and three-body decays. The branching fractions of the cocktail subcomponents are fixed from measurements when available~\cite{ATLAS_JpsiDs,LHCb-PAPER-2013-010}, and otherwise by analogy to $\decay{B^{0,\pm}}{D^* H_c X}$ decays. The ratio of two-body to quasi-two-body to three-body decay branching fractions is likelihood-constrained, with central values and uncertainties taken from summing the relative branching fractions of various modes measured in Refs.~\cite{LHCb-PAPER-2013-010,LHCb-PAPER-2016-055}. 
The semi-exclusive channel $\decay{B_c^+}{\jpsi D_s^{(*)+}}$, reconstructed in the $\decay{D_s^+}{K^+K^-\pi^+}$ final state, is used as a control sample to normalize the cocktail component. In this sample, the yield of $\decay{B_c^+}{\jpsi D_s^+}$ is obtained from a fit to the mass spectrum, while the $\decay{B_c^+}{\jpsi D_s^{*+}}$ contribution is inferred from the relative $\decay{B_c^+}{\jpsi D_s^{*+}}$ to $\decay{B_c^+}{\jpsi D_s^+}$ rate~\cite{PDG2024}. The corresponding yield for $\decay{D_s^+}{\mu^+ X}$ is then derived from the hadronic mode using the relevant branching fractions and reconstruction efficiencies, and included in the nominal fit as a constrained parameter.

To validate this modeling strategy, an independent cross-check evaluates the total $\decay{B_c^+}{\jpsi H_c X}$ background by extrapolating from inclusive $\decay{B_c^+}{\jpsi H_c X}$ candidates with hadronic \Dz, \Dp and \Dsp decays. The total yields obtained from this alternative approach are consistent with those derived from the nominal fit, confirming the robustness of the background estimation.

Three different sources of combinatorial background are considered. The largest component consists of random combinations of $\jpsi$ candidates originating from $\decay{B{^{+,0}}/B_s^0}{\jpsi X}$ decays with muon candidates from the rest of the event. This background is modeled with template histograms for the inclusive decays $\decay{B^0_{(s)}}{\jpsi X}$, where the unpaired muon originates from a different decay in the event. The $B^0$ channel simulation suffices to model both the $B^{0,+}$ decays. The relative contribution of the \Bs to \Bd decays is determined from the production cross sections~\cite{PDG2024}.

A different combinatorial background results from the pairing of uncorrelated muons to form a $\jpsi$ candidate. This background is modeled using $\jpsi$ candidates, which have an invariant mass above the $\jpsi$ mass window. The fitted yield of this background is likelihood-constrained around an estimate obtained by a fit to the \mup\mun invariant mass distribution using a double-sided Crystal Ball function to model the true $\jpsi$ candidates and an exponential function to model the combinatorial background. The width of the constraint is derived from the difference in estimations when a double Gaussian function is used as the signal model instead. 

Finally, another source of combinatorial background is from $\jpsi$ candidates originating from a $B_c^+$ decay that are combined with a muon from the semileptonic decay of an associated charm hadron produced in the fragmentation process. This background, hereafter referred to as the fragmentation background, is modeled using simulation.
A distinct feature of this background is the much shorter decay-time distribution relative to the other components in the fit, and it is reduced by restricting the fit to estimated decay times exceeding $0.45\ps$.  
To further reduce all combinatorial backgrounds, the angle between the momentum of the visible decay products and the flight direction of the $B_c^+$ candidate with respect to the PV is required to have a cosine greater than $0.9998$. Additionally, the $\jpsi$ and unpaired muon candidates must have a distance of closest approach of less than $0.05\mm$ when the vertex of the $\jpsi$ candidate is ahead of that of the $B_c^+$ candidate along the beam line and less than $0.13\mm$ when behind. 

The largest source of background, referred to as the misidentification (misID) background, originates from the combination of a $\jpsi$ originating from the decay of light $b$ hadrons with a charged hadron species $h^+ = (\pi^+, K^+, p)$  misidentified as a muon. The candidates from data used to construct this template pass the selection except for the identification requirements on the unpaired track, which is instead required to fail to leave enough hits in the muon system, leaving the sample enriched in various hadron species as well as electrons. To suppress the misID background, the component of the unpaired muon candidate momentum perpendicular to the $B_c^+$ candidate flight direction must be greater than $500 \mevc$. 

Weights representing the probability that a hadron with given kinematic properties is misidentified as a muon are computed from purified hadron samples~\cite{LHCb-PUB-2016-021}, while weights for fake tracks arising from unrelated hits are estimated using $\jpsi h^+$ simulation. The weights are applied to the $\jpsi h^+$ sample in order to generate the binned template representing the misID background component, with the overall normalization allowed to vary freely in the fit to the data. The procedure used to estimate the composition of the $\jpsi h^+$ sample is similar to that used in Refs.~\cite{LHCb-PAPER-2017-035,LHCb-PAPER-2022-039}. An alternative weighting procedure is developed based on methods used in Refs.~\cite{CDF:FullProbMethod, BcLifeTime-LHCb}. To mitigate model dependence, the $\mathcal{R}(\jpsi )$ value below is taken from the average of the two fit results using the default and alternative misID template. Half of the shift in the value of $\mathcal{R}(\jpsi)$ between the shapes is assigned as the systematic uncertainty due to misID composition estimation. An additional ``smearing" correction is applied, which accounts for the change in the momentum of the final-state muons relative to the original particles. The smearing correction was first developed in Ref.~\cite{LHCb-PAPER-2017-017} and uses simulated pions and kaons, which are identified as muons in order to derive the momentum correction. The fit is allowed to interpolate between templates with and without both the efficiency and smearing corrections. 

The fit model constructed from the sum of the templates representing the components described above is applied to the data using a binned maximum-likelihood method. All yields, except those described above, are allowed to vary in the fit.

The templates for the signal and normalization modes are derived from simulation, which uses the form factor model of Ref.~\cite{Ebert}. The simulation is weighted to match the form factors derived from fitting to LQCD synthetic data~\cite{Harrison:2025,Hammer}. The weights are allowed to vary in the fit within Gaussian constraints determined from the form factor uncertainties. 

Two BDT-based weighting procedures are applied to the simulation. The first provides a global correction for mismodeling of the underlying event and the resulting detector occupancy, using the overall event multiplicity and the number of tracks used in the PV reconstruction as inputs~\cite{XGBoosting}. The second corrects residual candidate-level differences in kinematic and vertexing distributions, using momentum and pseudorapidity variables, together with vertex-fit $\chi^2$, track DOCA with respect to the beam axis, impact-parameter $\chi^2$, and flight-distance $\chi^2$. The weights are derived in a normalization-enriched region (defined by $m_{\text{miss}}^2<1\,{\rm GeV}^{2}$ and $\tau<1.5\,{\rm ps}$) and applied to the entire fit region.


The ratio of the selection efficiencies between the signal and the normalization channels is determined from simulation to be $(45.4 \pm 0.6)\%$, where the uncertainty arises from the finite sizes of the simulated samples.

By using pseudoexperiments, small biases of $0.1\sigma_\mathrm{stat}$ or less were observed for the fit configurations using either misID template. These biases are corrected for in the final result.

\begin{figure}
    \centering
    \includegraphics[width=0.75\linewidth]{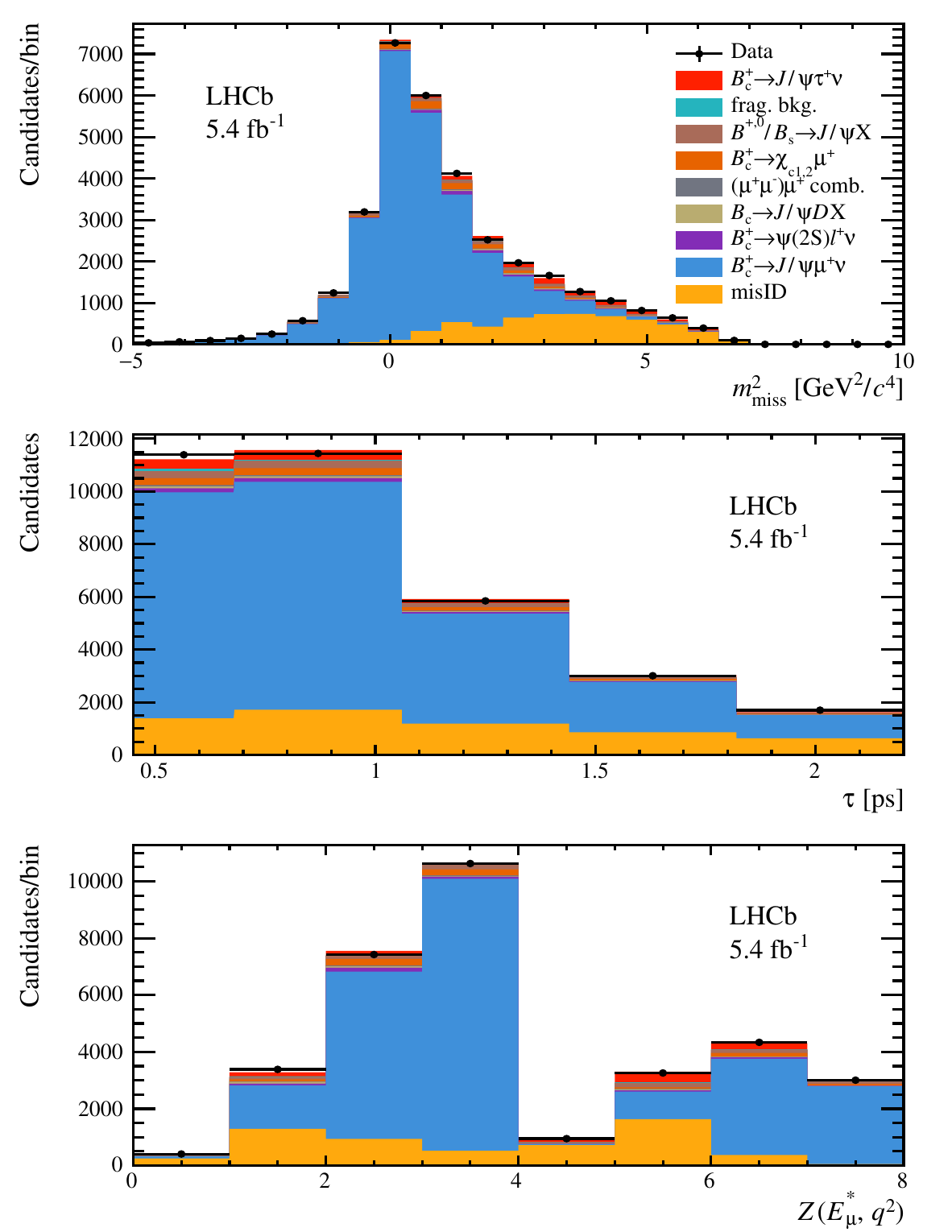}
    \caption{Distributions of (top) $m_{\text{miss}}^2$,  (middle) $B_c^+$ decay time $\tau$, and (bottom) $Z$ of the signal data with stacked projections of the fit model developed using the baseline fit strategy overlaid. Alternative visualizations and projections onto $q^2$ and $E_{\mu}^*$ can be found in the Supplemental Material \cite{ref:SuppMat}.
    }
    \label{fig:fit}
\end{figure}

The distributions of $m_{\text{miss}}^2$, $\tau$, and $Z$ in data with results of the fit to the sample after the final reweighting are shown in Fig.~\ref{fig:fit}. A total signal yield of $1.0\times 10^3$ is found, and $\mathcal{R}(\jpsi)$ is determined to be $0.52$ with a statistical uncertainty of $\pm 0.12$. The total systematic uncertainty is determined to be $7.9 \times 10^{-2}$, with the contributions of various effects listed in Table~\ref{tab:systematics} and described below.

\begin{table}
\begin{center}
\caption{Systematic uncertainties on the measured $\mathcal{R}(\jpsi)$ value, compared with the statistical uncertainty.}
\begin{tabular}{lr}
\hline
Source of uncertainty on $\mathcal{R}(\jpsi)$                     & Value ($\times10^{-2}$) \\
\hline
Simulation statistical uncertainty                                  & $5.9$  \\
$\decay{B_{c}^{+}}{\psi(2S) \lepton^+\nu}$ modeling                & $2.6$  \\
MisID composition estimation                                       & $2.5$    \\
Simulation corrections                                              & $2.0$   \\
$\decay{B_{c}^{+}}{\jpsi\,\ell^{+}\,\nu}$ form factors           & $1.8$  \\
$\decay{B_{c}^{+}}{\chi_{c} \,\mup \nu}$ scaling                  & $1.5$   \\
Efficiency ratio                                                    & $1.3$   \\
Fragmentation background modeling                                        & $1.3$   \\
MisID decay-in-flight correction                                   & $1.1$    \\
$(\jpsi \mup)$ combinatorial modeling                            & $0.6$   \\
MisID decay-in-flight smearing                                     & $0.3$  \\
$\decay{B_{c}^{+}}{\psi(2S) \,\ell^+ \nu}$ scaling                 & $0.3$   \\
$\mathcal{B}(\decay{\tau^+}{\mup \neum\overline{\nu}_{\tau}})$          & $0.2$   \\
Trimuon effect                                                      & $0.1$   \\
\hline
Systematic uncertainty                                              & $7.9$   \\
Statistical uncertainty                                             & $11.9$  \\
\hline
Total uncertainty                                                   &   $14.3$  \\
\hline
\end{tabular}
\label{tab:systematics}
\end{center}
\end{table}

The effect of the finite simulation sample size on the template shapes is determined using the Barlow--Beeston-lite procedure~\cite{Barlow:1993dm,Cranmer:2012sba}. The uncertainty from the $\decay{B_c^+}{\psitwos}$ form factors is evaluated by replacing the nominal templates based on Ref.~\cite{Ebert} with those based on Ref.~\cite{Kiselev}. The uncertainty associated with the final kinematic weighting is evaluated by retraining the BDT with the same inputs but alternative training parameters, and assigning half the difference in $\mathcal R(\jpsi)$ between the alternative and nominal fits.
    
In the baseline fit, LQCD results are used as Gaussian priors for the form factors with constraints taken from lattice uncertainties. To estimate the effect on $\mathcal{R}(\jpsi)$ due to the form factor constraints, an alternative fit is performed with the Gaussian constraints on the form factors removed, and half the shift in $\mathcal{R}(\jpsi)$ is taken as the systematic uncertainty. 
    
The uncertainty due to the contribution of the feed-down backgrounds $\decay{B_c^+}{\psitwos \ell^+ \nu_{\ell}}$ and $\decay{B_c^+}{\chi_{c(1,2)} \ell^+ \nu_{\ell}}$ are determined by half the shift in $\mathcal{R}(\jpsi)$ after fixing the yields from theoretical predictions.  A systematic uncertainty due to the potential mismodeling of the fragmentation background kinematic distribution is derived by performing the fit with an alternative template derived from the kinematic region containing the top 30\% of candidates sorted by the fraction of the charm hadron momentum to the $B_c^+$ momentum, and half the shift in $\mathcal{R}(\jpsi)$ relative to the baseline fit is taken as a systematic uncertainty. The uncertainty associated with the decay-in-flight efficiency correction is computed by using the efficiencies directly from the control samples and taking half the shift in $\mathcal{R}(\jpsi)$.
    
The modeling of the $\jpsi\mup$ combinatorial background gives an additional uncertainty. An exponential fit is performed to the $\jpsi\mup$ mass distribution in data above the $B_c^+$ mass region and compared to the simulated distribution in order to observe possible additional sources of combinatorial background unaccounted for by the considered $B_{d,s}$ decays. These include potential decays of $b$ baryons or the effect of unknown branching fractions. The $\jpsi \mup$ mass distribution in simulation is weighted to match the shape seen in data and half the resulting shift in $\mathcal{R}(\jpsi)$ from the fit relative to the nominal value, where no correction is applied, is taken as a systematic uncertainty. 
    
The systematic uncertainty of the misID decay-in-flight smearing procedure is calculated by taking half the shift in $\mathcal{R}(\jpsi)$ relative to an alternative fit where the pion and kaon momenta smearings are required to have the same ratio as observed in simulation rather than varying independently in the fit. A systematic uncertainty arises due to the presence of extra muons in the signal channel causing increased misID rates relative to those seen in the $\decay{\Dstarp}{\Dz (\to \Km\pip) \pip}$ calibration samples used for computing muon fake rates, known as the ``trimuon effect". This is evaluated by fitting using an angular selection that requires the muon tracks in the event to be sufficiently separated in the non-bending plane to avoid overlapping hits, and then taking half the shift in $\mathcal{R}(\jpsi)$. Finally, the uncertainty in the efficiency ratio and the branching fraction of $\decay{\tau^+}{\mup\neum \overline{\nu}_{\tau}}$ are propagated to $\mathcal{R}(\jpsi)$.

In conclusion, the ratio of branching fractions is measured to be
\begin{equation*}
    \begin{split}
        \mathcal{R}(\jpsi) &= \frac{\BF (\decay{B_c^+}{\jpsi \tau^+ \neut)}}{\BF (\decay{B_c^+}{\jpsi \mup \neum})} \\
        &= \FinalVal,    
    \end{split}
\end{equation*}
  which is consistent with the previous LHCb measurement based on the Run 1 data. The measurement represents a significant reduction of the systematic uncertainty as a result of the new selection criteria that led to the suppression of background processes, as well as the improved form factor information from LQCD. The result is within 1.8 standard deviations of the SM prediction from Ref.~\cite{Harrison:2025}. The significance of the $\Bcp\to\jpsi\taup\neut$ signal is estimated to be 3.6 standard deviations accounting for relevant systematic uncertainties. With the addition of this result, the world average value of $\mathcal{R}(\jpsi )$~\cite{HFLAV23} is in excess of the SM prediction by $2.4\sigma$.

\section*{Acknowledgements}
%
%
\noindent We express our gratitude to our colleagues in the CERN
accelerator departments for the excellent performance of the LHC. We
thank the technical and administrative staff at the LHCb
institutes.
We acknowledge support from CERN and from the national agencies:
ARC (Australia);
CAPES, CNPq, FAPERJ and FINEP (Brazil); 
MOST and NSFC (China); 
CNRS/IN2P3 and CEA (France);  
BMFTR, DFG and MPG (Germany);
INFN (Italy); 
NWO (Netherlands); 
MNiSW and NCN (Poland); 
MEC/IFA (Romania); 
MICIU and AEI (Spain);
SNSF and SER (Switzerland); 
NASU (Ukraine); 
STFC (United Kingdom); 
DOE NP and NSF (USA).
We acknowledge the computing resources that are provided by ARDC (Australia), 
CBPF (Brazil),
CERN, 
IHEP and LZU (China),
IN2P3 (France), 
KIT and DESY (Germany), 
INFN (Italy), 
SURF (Netherlands),
Polish WLCG (Poland),
IFIN-HH (Romania), 
PIC (Spain), CSCS (Switzerland), 
GridPP (United Kingdom),
and NSF (USA).  
We are indebted to the communities behind the multiple open-source
software packages on which we depend.
Individual groups or members have received support from
RTP (Australia), 
FWO Odysseus grant G0ASD25N (Belgium), 
Key Research Program of Frontier Sciences of CAS, CAS PIFI, CAS CCEPP (China); 
Minciencias (Colombia);
EPLANET, Marie Sk\l{}odowska-Curie Actions, ERC and NextGenerationEU (European Union);
A*MIDEX, ANR, IPhU and Labex P2IO, and R\'{e}gion Auvergne-Rh\^{o}ne-Alpes (France);
Alexander-von-Humboldt Foundation (Germany);
ICSC (Italy); 
Severo Ochoa and Mar\'ia de Maeztu Units of Excellence, GVA, XuntaGal, GENCAT, InTalent-Inditex and Prog.~Atracci\'on Talento CM (Spain);
the Leverhulme Trust, the Royal Society and UKRI (United Kingdom).





\addcontentsline{toc}{section}{References}
\bibliographystyle{LHCb/LHCb}
\bibliography{LHCb/main,LHCb/standard,LHCb/LHCb-PAPER,LHCb/LHCb-CONF,LHCb/LHCb-DP,LHCb/LHCb-TDR,LHCb/LHCb-PUB}

\newpage
\centerline
{\large\bf LHCb collaboration}
\begin
{flushleft}
\small
R.~Aaij$^{39}$\lhcborcid{0000-0003-0533-1952},
M.~Abdelfatah$^{71}$,
A.S.W.~Abdelmotteleb$^{59}$\lhcborcid{0000-0001-7905-0542},
C.~Abellan~Beteta$^{53}$\lhcborcid{0009-0009-0869-6798},
F.~Abudin\'en$^{61}$\lhcborcid{0000-0002-6737-3528},
T.~Ackernley$^{63}$\lhcborcid{0000-0002-5951-3498},
A.A.~Adefisoye$^{71}$\lhcborcid{0000-0003-2448-1550},
B.~Adeva$^{49}$\lhcborcid{0000-0001-9756-3712},
M.~Adinolfi$^{57}$\lhcborcid{0000-0002-1326-1264},
P.~Adlarson$^{87,44}$\lhcborcid{0000-0001-6280-3851}
% ...

{\footnotesize \it

% ...
$^{39}$Nikhef National Institute for Subatomic Physics, Amsterdam, Netherlands\\
% ...
$^{44}$National Center for Nuclear Research (NCBJ), Warsaw, Poland\\
% ...
$^{49}$Instituto Galego de F{\'\i}sica de Altas Enerx{\'\i}as (IGFAE), Universidade de Santiago de Compostela, Santiago de Compostela, Spain\\
% ...
$^{53}$Physik-Institut, Universit{\"a}t Z{\"u}rich, Z{\"u}rich, Switzerland\\
% ...
$^{57}$H.H. Wills Physics Laboratory, University of Bristol, Bristol, United Kingdom\\
% ...
$^{59}$Department of Physics, University of Warwick, Coventry, United Kingdom\\
% ...
$^{61}$School of Physics and Astronomy, University of Edinburgh, Edinburgh, United Kingdom\\
$^{62}$School of Physics and Astronomy, University of Glasgow, Glasgow, United Kingdom\\
$^{63}$Oliver Lodge Laboratory, University of Liverpool, Liverpool, United Kingdom\\
% ...
$^{71}$Syracuse University, Syracuse, NY, United States\\
% ...
$^{87}$Department of Physics and Astronomy, Uppsala University, Uppsala, Sweden, associated to $^{62}$\\
% ...
}
\end{flushleft}




\end{document}